\documentclass[twocolumn,showpacs, preprintnumbers, amsmath, amssymb]{revtex4}
\usepackage{dcolumn}
\usepackage{graphicx}
\usepackage{bm}
\usepackage{subfigure}
\begin{document}
\title{The accuracy of roughness exponent measurement methods}
\author{Jan {\O}ystein Haavig Bakke}
\email{Jan.Bakke@ntnu.no}
\author{Alex Hansen}
\email{Alex.Hansen@ntnu.no} 
\affiliation{Department of Physics, Norwegian University of Science and
Technology, N--7491 Trondheim, Norway}
\date{\today}
\begin{abstract}
We test methods for measuring and characterizing rough profiles with
emphasis on measurements of the self-affine roughness exponent, and
describes a simple test to separate between roughness exponents originating
from long range correlations in the sign signs of the profile, and roughness
exponents originating from L{\'e}vy distributions of jumps. Based on tests on
profiles with known roughness exponents we find that the power spectrum
density analysis and the averaged wavelet coefficients method give the best
estimates for roughness exponents in the range $0.1$ to $0.9$. The error-bars
are found to be less than $0.03$ for profile lengths larger than $256$, and
there are no systematic bias in the estimates. We present quantitative 
estimates of the error-bars and the systematic error and their dependence on 
the value of the  roughness exponent and the profile length. We also quantify 
how power-law noise can modify the measured roughness exponent for measurement 
methods different from the power spectrum density analysis and the second order
correlation function method.
\end{abstract}
\maketitle

\section{Introduction}
\label{sec:intro}
In 1990 Bouchaud et al.\ \cite{blp90} proposed that the roughness exponent for 
profiles from three dimensional fracture surfaces was universal - independent 
of material properties and fracture mode. This has since been the working 
hypothesis in the community studying fracture roughness. To claim universality 
one needs good measurements and to be aware of the inherent biases and 
limitations of the different methods used for measuring the roughness 
exponent. The use of several independent methods are a prerequisite for a good 
estimate of the roughness exponents. This is even more true now as the study 
of self-affine surfaces goes beyond simple measurements of the roughness 
exponent and the higher order statistics and the shape of the 
height-difference distribution function $p(\Delta h,l)$ are also 
studied. \cite{bpsv06,smdmhbsvr07} Now a thorough measurement of the 
self-affinity of a surface should  also include checks for corrections to 
scaling and a survey of the higher order statistics i.e. multi-scaling, and if 
needed a check for anomalous scaling.

\begin{figure}[tbp]
  \begin{center}
\includegraphics[scale = 0.3]{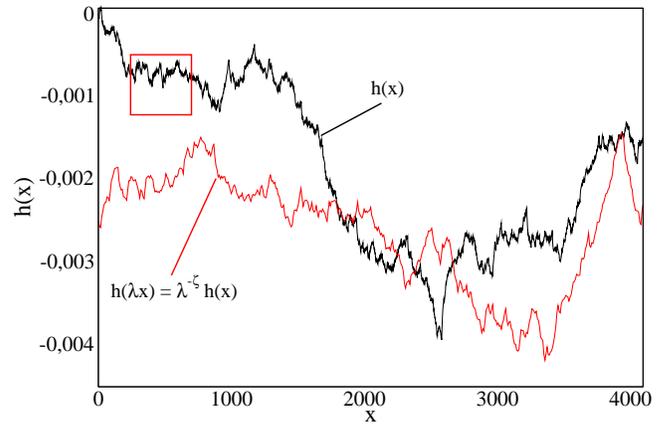}
\caption{\label{fig:sascaling}(Color on-line) If a surface of a given linear 
size $L_1$ is with $L_2 = \lambda L_1$ in the horizontal direction,
the surface must be scaled with $h(\lambda x) = \lambda^{-\zeta}h(x)$ in the 
vertical direction to be statistically similar. $\zeta$ is the roughness, or 
Hurst exponent, measuring the degree of anisotropy.  $\zeta = 1$ gives 
self-similar scaling. In this figure the part of the black profile which is 
inside the square is rescaled with the scaling relation above.}
\end{center}
\end{figure}

Research on surfaces morphology with focus on the scaling properties of the 
surfaces has been pursued since the work of Mandelbrot et al. in 
1984. \cite{mpp84} These studies have not been restricted to fracture surfaces.
Examples of surfaces that have been studied during the last twenty years are 
fluid fronts in disordered media, fire fronts in paper, atomic deposition 
surfaces, fracture surfaces, DNA base-pair sequences and the time signal of the
heart rhythm. These surfaces have been shown to have statistically self-affine 
scaling properties. The self-affine scaling is an anisotropic scaling of the 
system. This is seen in the scaling of the height-difference distribution 
function
\begin{equation}
  p(\lambda^{\zeta}\Delta h,l) \propto \lambda^{-\zeta}p(\Delta h,l)
  \label{eq:pdhscaling}
\end{equation}
An example of self-affine scaling in two dimensions is shown in 
Fig. \ref{fig:sascaling}.
Systems with a common roughness exponent are said to belong to the same 
universality class, and are therefore controlled by the same fundamental 
physical law. An early summary of different universality classes and the
different surfaces studied can be found in Barab{\'a}si and 
Stanley \cite{bs95}. 

The surfaces described above are constructed in different ways. 
Restricting the discussion to two-dimensional surfaces or profiles
we can divide the profiles into three groups. The first group of profiles 
grow from an initial planar profile or line into a rough profile.
Examples from this group of  profiles are wetting fronts, deposition fronts 
and three dimensional fracture surfaces constrained to grow in a plane.
The second group of profiles grow at the end points. Examples from this group
of profiles are DNA base-pair sequences, any times series and fractures growing
in two-dimensional systems. The third group is made by coalescence of micro 
cracks. Examples from this group of cracks is crack growth by plastic micro 
void coalescence. The initially flat profiles are often characterized by the 
Family-Vicsek scaling relation of the profile width  
$w(L,t) \propto L^{\alpha}f(t/L^z)$ \cite{fv85},where $\alpha$ is the roughness
exponent measured when the system is saturated and $z = \alpha/\beta$ is the 
dynamic exponent. $\beta$ is the growth exponent characterizing the growth of 
the profile width before saturation $w(L,t) \propto t^{\beta}$ The values of 
the three exponents then give the universality class. Profiles that do not obey
Family-Vicsek scaling are said to have anomalous scaling.

The concept of anomalous scaling can not be applied to the second group of 
profiles as they are only $1+1$ dimensional, while the profiles in the first 
group are $2+1$ dimensional. The fracture profile in two dimensions are 
strictly speaking not $1+1$ dimensional since there are two space dimensions 
and no time dimension. Each single part of the fracture is frozen once it is 
placed in the system. It is therefore no time evolution of the fracture in 
$2+1$ dimensions. 

The aim of this paper is to give a survey of the different methods used for 
measuring the roughness exponent and give an estimate of the expected errors 
and biases. This paper will use some of the methods used by Schmittbuhl et 
al.\ in \cite{svr95}, excluding the fractal measurements and the return methods
since these methods have already been tested there. We repeat the test done 
with the local window methods and the power spectrum density analysis. Other 
methods such as the detrended fluctuation analysis \cite{dfa}, the averaged 
wavelet coefficient method\cite{shn98} and the height difference distribution 
are also included.  To accurately asses the results from the different methods 
the methods must be applied to profiles with known roughness exponent. Both 
the Voss method \cite{voss85} and the wavelet method \cite{sh02} for generating
profiles have been used and analyzed. We will also study the systematic error 
introduced by adding a power law noise to the signal, and discuss that both 
long range correlations in the sign change and a power law noise can give 
surfaces with roughness exponents. This is motivated by recent results for the 
central force and fuse model of fracture.\cite{brh07,bh07}. Our interest in the
phenomena of self-affine surfaces comes from the studies of fracture surfaces 
both experimental and numerical, and the difficulties we have encountered 
while measuring the roughness exponent.

The structure of this paper is as follows: In Sec.\ \ref{sec:meas} the 
different measurement methods are presented. In Sec.\  \ref{sec:acc} we 
presents the results from the measurements done on the generated profiles. 
Samples of the power law scaling for each method is presented and compared with
the other methods. The results for both the generation methods described in the
last paragraph are shown. We find that the power spectrum density analysis and 
the averaged wavelet coefficient method give the most accurate for roughness 
exponents in the range $0.1-0.9$. In Sec.\ \ref{sec:sizedep}  we study the size
dependence on the systematic errors for the different methods. We observe as 
expected that the errors decreases for larger system sizes, but several of the 
measuring methods have relatively large systematic error and the size of this
error depends on the measured roughness exponent. In Sec.\ \ref{sec:powlaw} we 
show that power law noise might disguise as self-affinity when profiles are 
analyzed with some methods. We then describe how one can separate the 
contribution from the power law noise and the sign change correlation to the 
roughness exponent. And finally in Sec.\ \ref{sec:disc} we give a summary and
come with some recommendations to researchers studying surface roughness.

\section{Measurements of the roughness exponent}
\label{sec:meas}
Until recently one used different methods for measuring the roughness 
exponent on experimental and numerically made fracture surfaces. On surfaces 
from experiments one applied local (or intrinsic) methods which use scaling 
with a (intrinsic) length scale $l$ much smaller than the system size $L$ or 
spectral methods like the power spectrum density. On the other hand, for 
surfaces from numerical simulations one applied global (or extrinsic) methods 
which use scaling with the system size. This was was done by necessity as in 
experiments it is not easy to make samples of the same material with sizes 
ranging over several orders of magnitude, or experimental setups that can do 
measurements over the same orders of magnitude. While for numerical simulations
the restriction in computing power made it difficult to create large enough 
samples to use the local methods. But during the last few years and due to the 
growth in computing power numerical simulations of fracture have produced large
enough samples in such numbers that both the real space and spectral local 
methods are used in numerical studies of fracture surfaces.

Below we will describe some local and global methods which are used today for
measuring the roughness exponent. All the methods we consider in this paper 
work on profiles. I.\ e. profiles in two-dimensional planes cut  from a surface
embedded in three dimensions or a one-dimensional trace or path embedded in 
two-dimensions.

The local window methods all measure the scaling of a characteristic width as a
function of the window size. The width is a measure of how large the 
fluctuations of the profiles in a window of length $l$ are. We will look at 
three different methods which use different definitions of the characteristic 
width: 1. The variable bandwidth method (VB)
\begin{equation}
  w_{VB}(l) = (\langle (h(x) - \bar{h})^2\rangle_L)^{1/2} \propto l^{\zeta}\mbox{,}
\label{eq:vb}
\end{equation}
where $\bar{h} = \langle h(x) \rangle$. 2. The detrended fluctuation analysis 
(DFA), where the local linear trend in each window is subtracted
\begin{equation}
  w_{DFA}(l) = (\langle (h'(x) - \bar{h'})^2\rangle_L)^{1/2} \propto l^{\zeta}\mbox{.}
\label{eq:lwd}
\end{equation}
3. the maximum - minimum method (MM)
\begin{eqnarray}
  w_{MM}(l) & = &\langle \mbox{max}(h(x),x \in \{x_o,x_o+l\} \nonumber\\ 
  & & - \mbox{min}(h(x),x \in \{x_o,x_o+l\}\rangle_L \propto l^{\zeta}.
\label{eq:lwminmax}
\end{eqnarray}
$\langle\cdots \rangle_L$ means averaging over the system size $L$.  Averaging 
over different samples is implied.  The roughness exponent $\zeta$ is then 
found as the power law scaling with $l$ of the characteristic width.

An alternative measuring method is to look at the second order correlation
function
\begin{equation}
  C_2(r) = \langle (h(x-r)-h(x))^2\rangle_L^{1/2} \propto r^{\zeta};
  \label{eq:c2}
\end{equation}

The roughness exponent can also be found from the two spectral methods as the 
scaling of the power spectrum $P(k)$, \cite{falconer90} which is the Fourier 
transform of the correlation function Eq. (\ref{eq:c2}), with the wave number 
for the power spectrum analysis
\begin{equation}
  P(k) \propto k^{-2(\zeta + 1)}\mbox{,}
\label{eq:psd}
\end{equation}
and as the scaling of the averaged wavelets coefficients $W[h](a)$ as a 
function of the scale variable $a$
\begin{equation}
  W[h](a) \propto a^{+\zeta + 1/2}.
\label{eq:awc}
\end{equation}

The global average width method is similar to the intrinsic one, but with the 
window size $l$ substituted with the system size $L$.
\begin{equation}
  W(L) = \langle h(x) - \bar{h}\rangle \propto L^{\zeta}\mbox{,}
  \label{eq:lW}
\end{equation}
Similarly we have the global maximum-minimum method
\begin{eqnarray}
  W_{MM}(L) & = & \langle \mbox{max}(h(x),x \in \{0,L\} \nonumber\\
  & & - \mbox{min}(h(x),x \in \{0,L\}\rangle \propto L^{\zeta}.
\label{eq:lWminmax}
\end{eqnarray} 

The roughness exponent measured by the global methods give the same results as 
the local methods only of the scaling of $W(L)$ is the same as the scaling of 
$w(l)$, $l \ll L$. One example of when this is not the case is when the 
profiles shows anomalous scaling. \cite{lr96} In this paper we will not 
consider the global methods.

In two recent papers the question of whether the fracture surfaces measured
in experiments are self-affine or multi-affine has been discussed.
\cite{bpsv06,smdmhbsvr07} In Santucci et al.\ \cite{smdmhbsvr07} two slightly 
different methods for measuring the fracture roughness have been discussed. 
The basis for these methods is to assume that the distribution function 
$p(\Delta h,l)$, $\Delta h(l) = h(x+l)-h(x)$ is 'Gaussian-like'. The first 
method was introduced in the studies of directed polymers in random 
media. \cite{hh44}. The $k$-th moment of $h(x)$. is defined below
\begin{equation}
  C_k(l) = \langle|h(x+l)-h(x)|^k\rangle^{1/k}
  \label{eq:gk}
\end{equation}
To check whether or not the surface have self-affine or multi-affine scaling 
one calculates the ratio of the $k$-th to the second moment 
\begin{equation}
  R_k(l) = \frac{\langle |h(x+l)-h(x)|^k\rangle^{1/k}}{\langle(h(x+l)-h(x))^2\rangle^{1/2}}
  \label{eq:rk}
\end{equation}
which for a true Gaussian distribution should reduce to
\begin{equation}
  R_K^G(l) = R_K^G = \sqrt{2} \left( \Gamma ( \frac{ \left( \frac{k+1}{2}) \right) }{\sqrt{\pi}} \right)^{1/k}.
  \label{eq:rkg}
\end{equation}

By plotting $R_k/R_k^G$ one should then get straight lines if the  surface is 
self-affine since $\zeta_k = k\zeta$ when the surface is self-affine. If these 
lines should fall on top of each other, the underlying distribution is also
Gaussian. The roughness exponent can be found by finding the slope of 
$C_k/C_k^G$, where $C_K^G$ is the $k$-th moment for a true Gaussian 
distribution, in a double logarithmic plot, which will return the same 
roughness exponent as $C_2(r)$.

The other method is to construct the distributions $p(\Delta h,l)$ of the 
height difference over distance $l$ and plotting 
$p(\Delta h,l)\sqrt{2\pi \sigma^2}$  versus $(\Delta h)/\sqrt{2\sigma^2}$, 
where $\sigma^2$ is the fluctuations of $\Delta h$ over $l$.  For Gaussian 
distributions the plot should be a parabola pointing downward in a 
semi-logarithmic plot. The roughness exponent can then be found from the power 
law scaling of slope of the fluctuations in 
$\Delta h$, $\sigma(l) \propto l^{\zeta}$.

Many of the methods described above involve the scaling of a characteristic 
width with a window length. At least two mechanisms can lead to the increase in
the characteristic width.\cite{falconer90} The first one is that there are 
spatial correlations in the sign change of the steps in the profile, and the 
second one is a L{\'e}vy like jump distribution. To check how these two 
mechanisms contribute to the measured effective roughness exponent on can do 
two different modifications to the measured profiles. If one sets the size of 
each jump equal to unity using
\begin{equation}
  h_0(x) = \lim_{q \rightarrow 0} \int_0^L \mbox{sgn}(h(x))|h(x)|^q dx\mbox{,}
  \label{eq:h0}
\end{equation}
one will only measure the characteristic width that are caused by the 
correlations in the sign changes as any information carried by the amplitude 
will be removed. It has been shown \cite{hm06} that for $h_0(x)$ the 
roughness exponent is
\begin{equation}
  \zeta = \max(\frac{1}{2},\zeta_h)
  \label{eq:zeta0}
\end{equation}
where $\zeta_h$ is the roughness exponent of $h(x)$. Thus for a roughness 
exponent less $1/2$ some information in the roughness exponent will be in the 
jump distribution.

To remove any correlation that might be in the sign changes, but keep the 
information in the jump distribution one can randomly rearrange the position 
for each jump. The characteristic width measured on this randomly rearrange 
profiles $h_r(x)$ will now only depend on the jump distribution, and therefore
the measured roughness exponent is due to the jumps.

\section{Accuracy of the measurement methods}
\label{sec:acc}
In Schmittbuhl et al.\ \cite{svr95} the reliability of self-affine measurements
is addressed for some of the methods above. We will repeat the measurements for
the power spectrum density analysis and local window methods, and compare with 
the additional methods from Sec.\ \ref{sec:meas} to check the validity of the 
roughness exponents we measure from the generated self-affine fracture 
surfaces.

We base our choice of good measurement methods on two criteria: 1. The fitting 
of the power law should be good i.e. good linearity in the double logarithmic 
plots. 2. The systematic error for the method should be small and stable over 
a range of different roughnesses.

\begin{center}
  \begin{figure}[tbp!]
    \includegraphics[scale = 1.0]{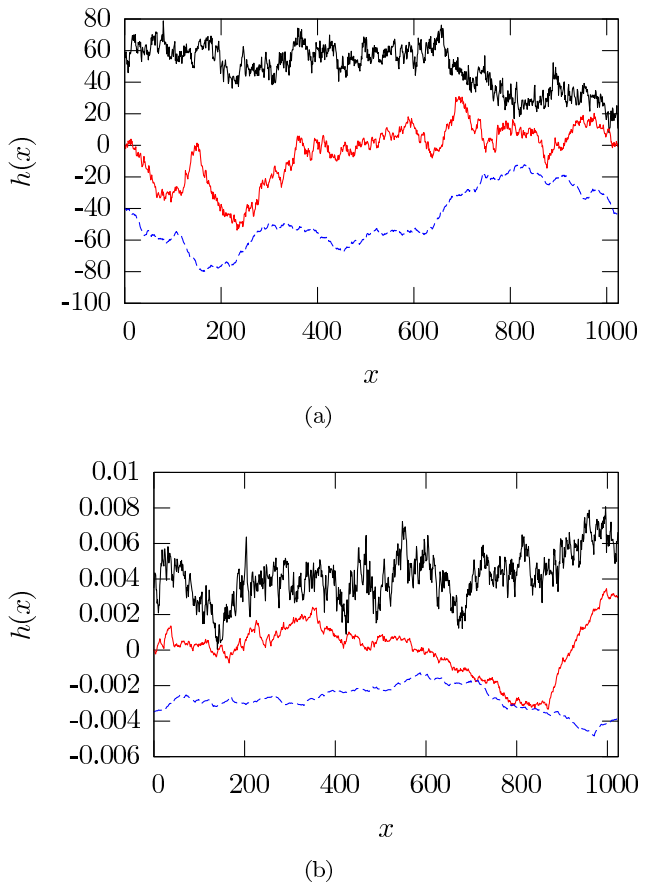}
    \caption{\label{fig:bothprofiles}(Color on-line) Sample profiles from the 
Voss algorithm (a) and the wavelet algorithm (b). The profiles are from top to 
bottom for roughnesses $0.2$, $0.5$ and $0.8$. To make the scale of the 
fluctuations the same for samples with different $\zeta$ the samples are 
rescaled with $W_{\zeta}(L)/W_{0.5}(L)$.}
  \end{figure}
\end{center}

We generate artificial surfaces using two methods. The first method is the
Voss method \cite{voss85}, and the second is the wavelet method \cite{sh02}.
The profiles are generated by first creating long profiles of a length much
larger than the profile sizes we want to study.  We then cut out a piece of 
desired length at a random position from the long profiles. In this study the 
long profiles had a length of 65536. Before we measure the roughness exponent
we remove any linear trend in profiles. The tests are done for 
$\zeta \in \{0.1,0.9\}$ and system sizes $L \leq 16384$. For each roughness 
exponent and each system size $100$ samples were generated. Samples of these
profiles are shown in Fig.\ \ref{fig:bothprofiles}. Before one starts to
measure the roughness exponent it is from our experience always a good idea
to visually inspect the profiles and compare them to self-affine profiles
with known roughness exponents.

\begin{center}
  \begin{figure}[tbp!]
    \includegraphics[scale = 1.0]{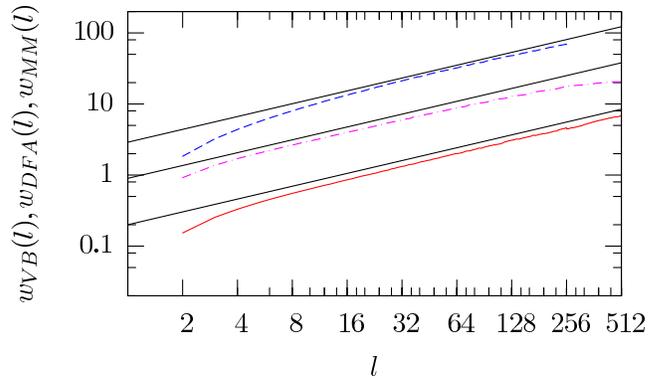}
    \caption{\label{fig:lwmethods}(Color on-line) Measurements of the roughness
exponent using intrinsic window methods. From the top: The local MAX-MIN method
Eq. (\ref{eq:lwminmax}), the variable bandwidth method Eq. (\ref{eq:vb}) and 
the detrended fluctuation analysis Eq. \ref{eq:lwd}.}
  \end{figure}
\end{center}

\begin{center}
  \begin{figure}[htbp!]
    \includegraphics[scale = 1.0]{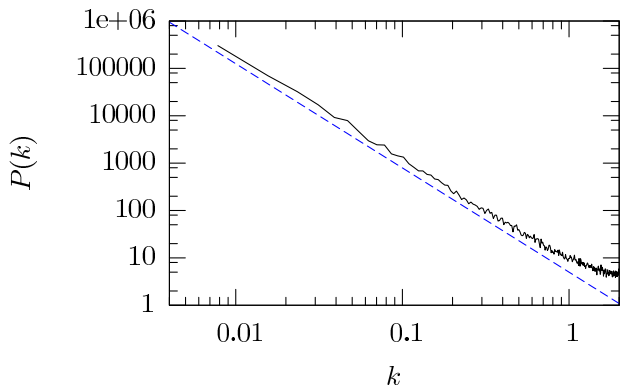}
    \caption{\label{fig:psdmethod}(Color on-line) Measurements of the 
roughness exponent using the power spectrum density analysis, 
Eq. \ref{eq:psd}.}
  \end{figure}
\end{center}

\begin{center}
  \begin{figure}[tbp!]
    \includegraphics[scale = 1.0]{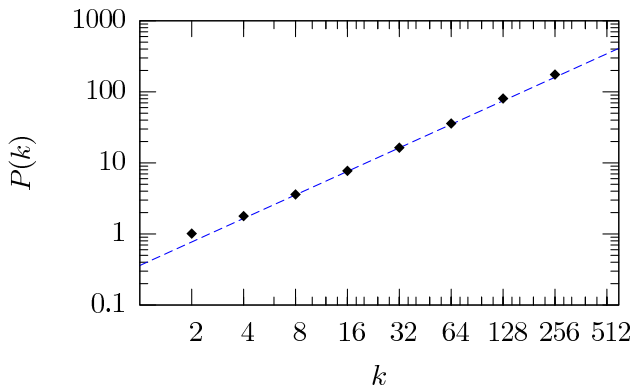}
    \caption{\label{fig:awcmethod}(Color on-line) Measurements of the 
roughness exponent using the averaged wavelet coefficients method 
Eq. \ref{eq:awc}.}
  \end{figure}
\end{center}

\begin{center}
  \begin{figure}[tbp!]
    \includegraphics[scale = 1.0]{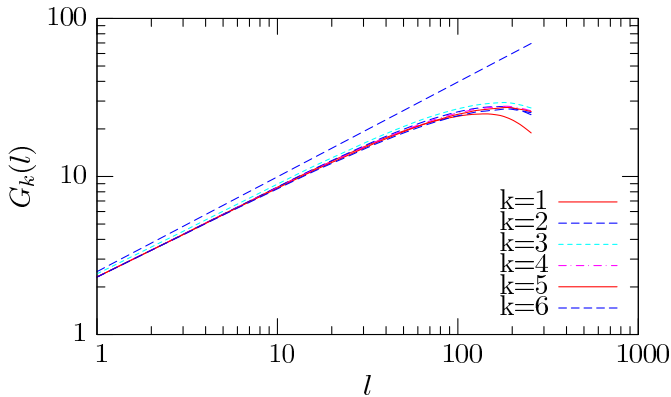}
    \caption{\label{fig:gka}(Color on-line) Measurements of the 
roughness exponent using second order correlation function Eq. (\ref{eq:c2}).}
  \end{figure}
\end{center}

\begin{center}
  \begin{figure}[tbp!]
    \includegraphics[scale = 1.0]{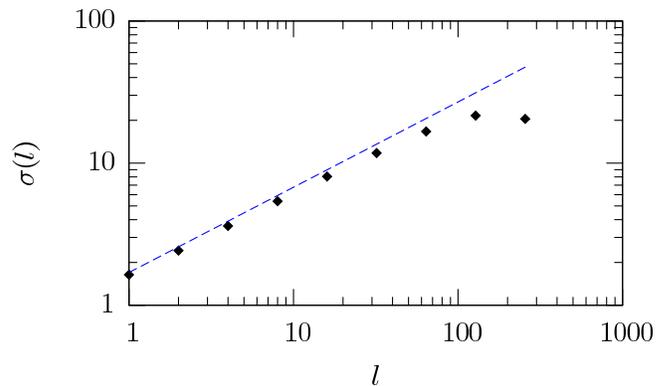}
    \caption{\label{fig:psigmamethods}(Color on-line) Measurements of the 
roughness exponent using the scaling of $\sigma(l)$ for the distribution 
$p(\Delta H,l)$}
  \end{figure}
\end{center}

\begin{center}
  \begin{figure}[tbp!]
    \includegraphics[scale = 1.0]{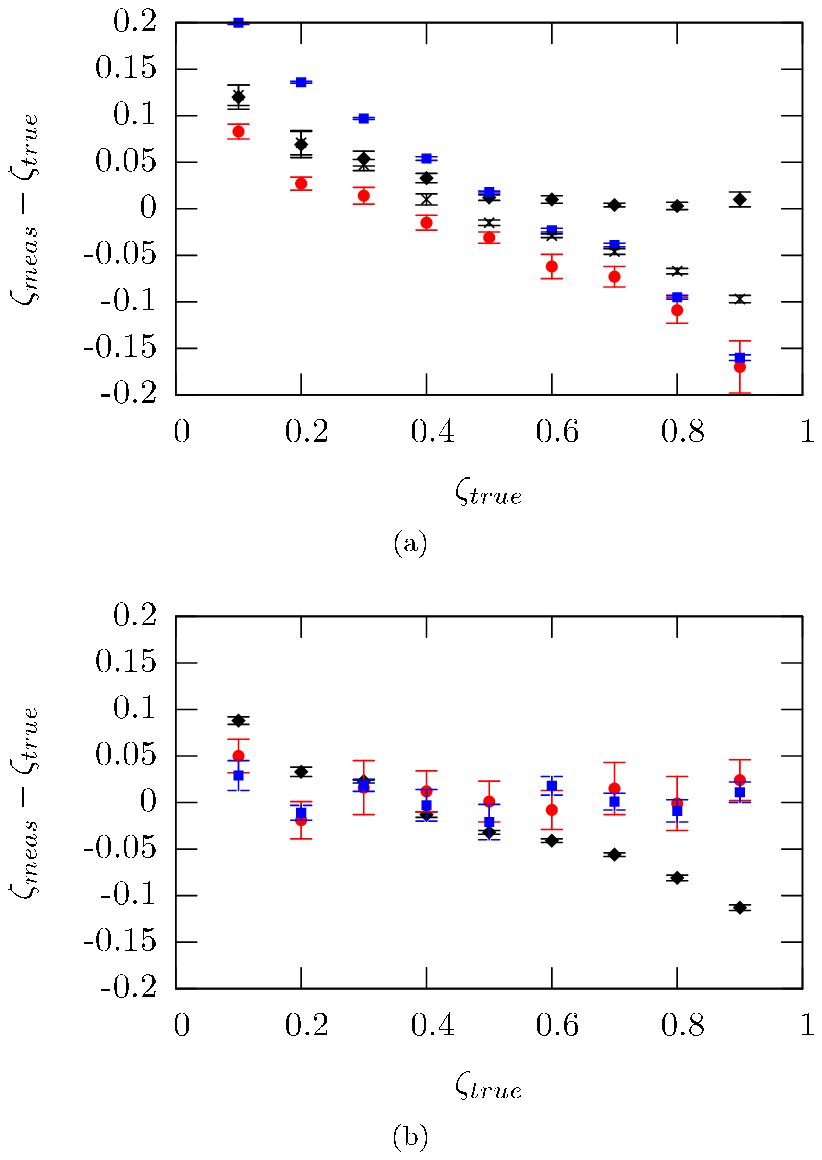}
   \caption{\label{fig:voss}(Color on-line) $\zeta_{measured} - \zeta_{true}$
for profiles generated by the Voss algorithm. a) The detrended fluctuation 
analysis ($\blacklozenge$), the variable bandwidth ($\bullet$), the Max-Min 
($\blacksquare$) and the standard deviation of $\Delta h(l)$ ($+$). b) Second order correlation function ($\blacklozenge$), power spectrum density analysis 
($\bullet$) and averaged wavelet coefficients ($\blacksquare$).}
   \end{figure}
\end{center}

\begin{center}
  \begin{figure}[tbp!]
    \includegraphics[scale = 1.0]{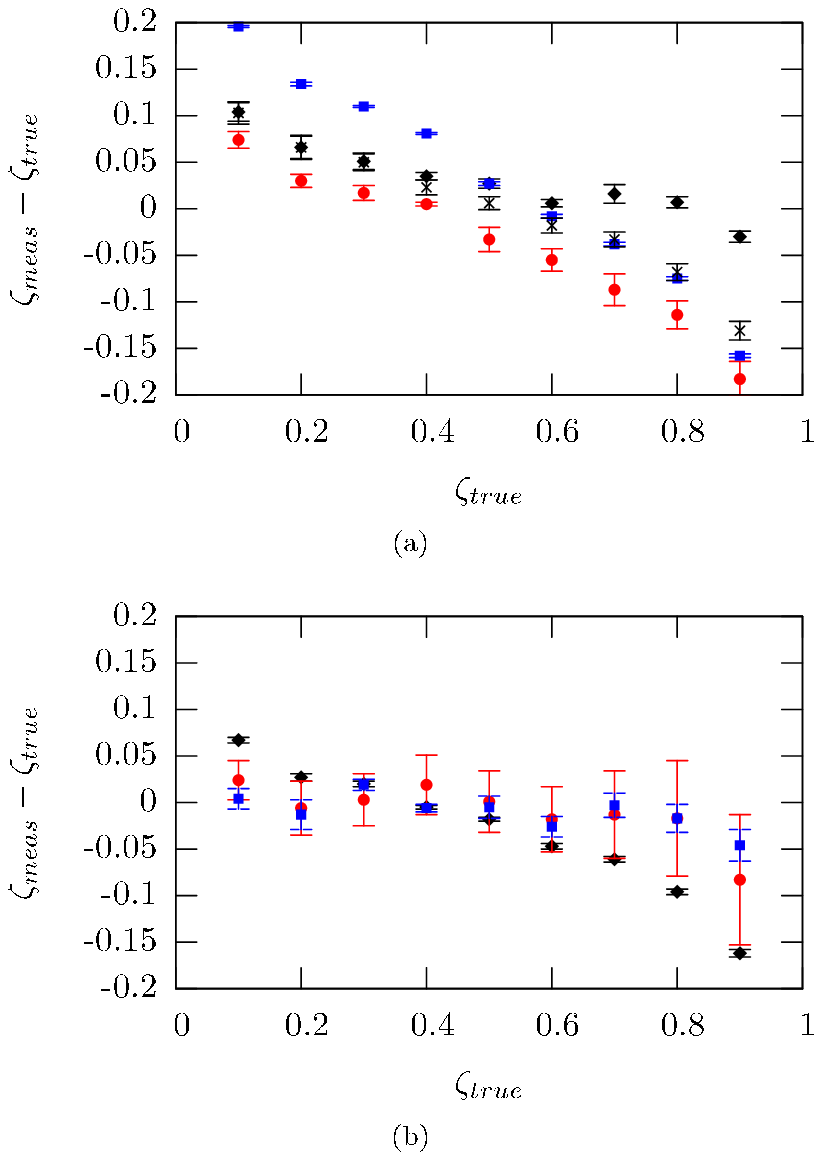}
   \caption{\label{fig:wavelet}(Color on-line) 
$\zeta_{measured} - \zeta_{true}$ for profiles generated by the wavelet 
algorithm. a) The detrended fluctuation analysis ($\blacklozenge$), the 
variable bandwidth ($\bullet$), the Max-Min ($\blacksquare$) and the standard 
deviation of $\Delta h(l)$ ($+$). b) Second order correlation function 
($\blacklozenge$), power spectrum density analysis ($\bullet$) and averaged 
wavelet coefficients ($\blacksquare$).}
   \end{figure}
\end{center}

\begin{center}
  \begin{figure}[tbp!]
    \includegraphics[scale = 1.0]{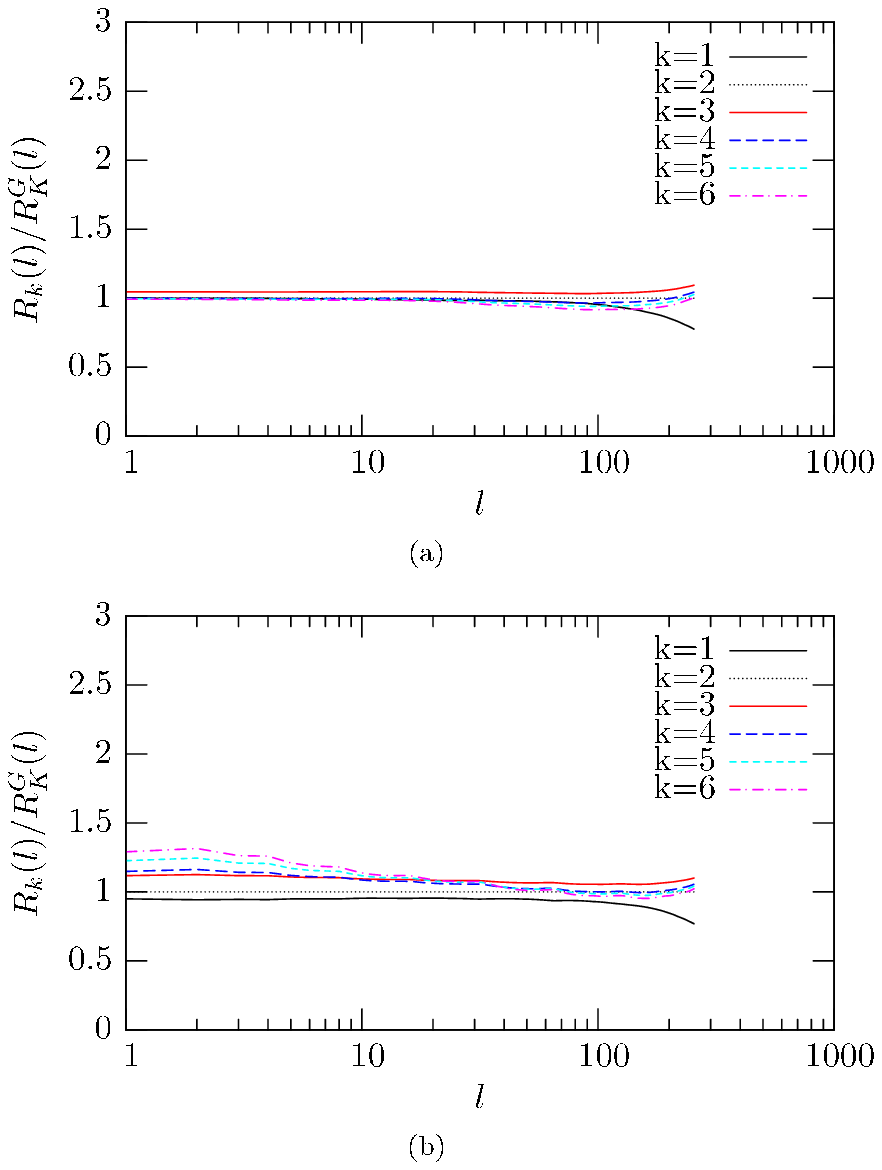}
   \caption{\label{fig:rka}(Color on-line) $R_k/R_k^G$ for profiles of length 
512 made by a) the Voss algorithm and b) the wavelet method for $\zeta = 0.7$.}
   \end{figure}
\end{center}

\begin{center}
  \begin{figure}[tbp!]
    \includegraphics[scale = 1.0]{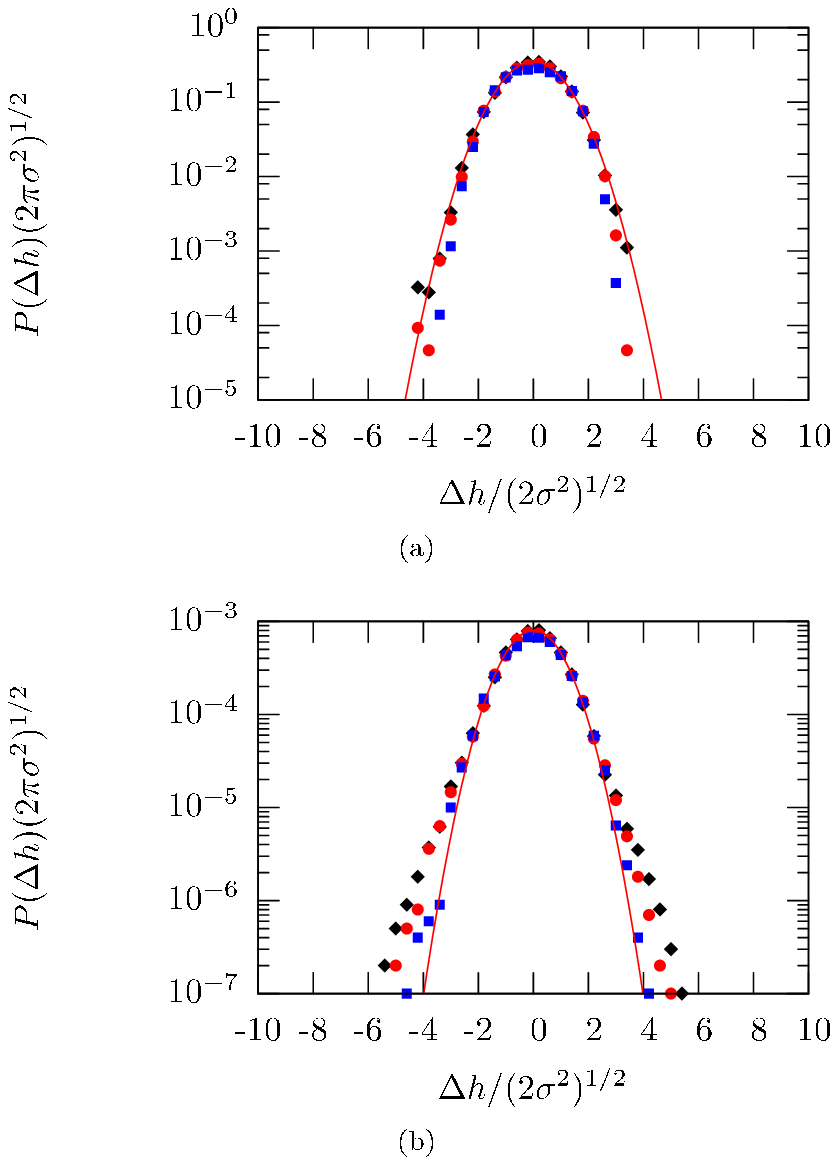}
    \caption{\label{fig:pdh}(Color on-line) $p(\Delta h,l)$ for profiles of 
length $512$ made by a) the Voss algorithm and b) the wavelet method for 
$\zeta = 0.7$ Three different length scales are presented here: $l = 64$ 
($\blacklozenge$), $l = 128$ ($\bullet$) and $l = 256$ ($\blacksquare$).}
   \end{figure}
\end{center}

In Figs. \ref{fig:lwmethods} to \ref{fig:psigmamethods} we present the 
different measurements for $L=512$. This system size is now numerically 
accessible for many different numerical models of fracture and large enough to 
give good estimates for the roughness exponent using the local methods. The 
profiles were made with $\zeta = 0.6$ and all straight lines in the figures 
represents this $\zeta$-value. These plots are from the Voss profiles. From 
Fig. \ref{fig:lwmethods} one see that for the local window methods in 
Eq.\ (\ref{eq:vb}) to Eq.\ (\ref{eq:lwminmax}) the detrended fluctuation 
analysis gives the best estimate of the roughness exponent. The detrended 
fluctuation analysis gives a straight line in the double logarithmic plot 
from $l = 16$ to $l=512$ while the variable bandwidth and the Max-Min data are 
more curved compared to the $\zeta = 0.6$ lines.
The power spectrum density method in Fig. \ref{fig:psdmethod} as well as the 
averaged wavelet coefficients method in Fig. \ref{fig:awcmethod} both give 
good estimates of the roughness exponent. For $C_k/C_k^G$ we can see in 
Fig. \ref{fig:gka} that the second order correlation function gives an 
estimate of the roughness exponent comparable which is below the correct value.
In addition we see that the collapse of the different moments is showing that 
the profiles are self-affine, not multi-affine. Fig. \ref{fig:psigmamethods} 
shows the scaling of $\sigma(l)$ which give results similar to the second 
order correlation function.
When we compared the result for the profiles made with the Voss algorithm in 
Fig. \ref{fig:voss} and the profiles made with the wavelet method in 
Fig. \ref{fig:wavelet} we notice some differences. Using the power spectrum 
density analysis and the averaged wavelet coefficients methods the profiles 
made by both methods we obtain good measurements of the roughness exponent. 
The results for when using the averaged wavelet coefficient method on the 
wavelet generated profiles were no surprise as this should a priori give 
correct results. When one use the detrended fluctuation analysis and the 
variable bandwidth, the Max-Min and $C_2$ methods we see a more pronounced 
deviation from the true roughness exponent used to create the profiles for the 
highest $\zeta$ for the wavelet generated profiles. 
This result suggests that the profiles made by the wavelet method is not as 
good as the Voss methods for generating self-affine profiles. A test of the 
self-affinity of the profiles confirms this. In Fig. \ref{fig:rka} we clearly 
see that the wavelet generated profiles have corrections to scaling at small
scales. This correction to scaling can also be seen for $p(\Delta h,l)$ in 
Fig. \ref{fig:pdh}. For the high $\zeta$-values the wavelet generated profiles 
the tail is broader than a Gaussian distribution. This effect becomes larger 
for the largest $\zeta$s. The profiles generated by the Voss algorithm is more 
narrow than the Gaussian distribution, and also for these profiles is the 
effect greatest for the largest $\zeta$-values. In the rest of this paper we 
will only consider the profiles made by the Voss method.
From Figs.\ \ref{fig:voss} and \ref{fig:wavelet} one can see that the
power spectrum density analysis and the averaged wavelet coefficients method 
give the best estimates for the roughness exponent in the range $0.1 - 0.9$.
For these two methods the test measurements done here gave error bars from the 
power law fits equal or less than $0.03$. The detrended fluctuation analysis 
gave good estimates above $\zeta = 0.5$, and systematically to high estimates 
for  $\zeta < 0.5$. The other local methods did only give  good estimates 
around $\zeta = 0.5$, and showed an systematic drift towards lower $\zeta$ 
values for $\zeta > 0.5$, and towards higher $\zeta$ values for $\zeta < 0.5$. 
Of the global $W(L)$ gave the most accurate estimates, but gave systematically 
to high estimates for $\zeta < 0.5$.

The results for the roughness exponent were done with the following limits
on for the regions where we did a least square fit. For the variable bandwidth 
method, the Max-Min method and detrended fluctuation analysis the regression
was done in the range $l \in [16,L/2]$. For the second order correlation
function the regression was done for $l \in [2,L/8]$. For the power spectrum 
analysis the regression was done for $k \in [0,0.2]$. For the averaged 
wavelet coefficients method the regression was done for $a \in [8,L/4]$. For
the standard deviation of the height-difference the regression was done for
$l \in [1,L/16]$. 

\begin{center}
  \begin{figure}[tbp!]
    \includegraphics[scale = 1.0]{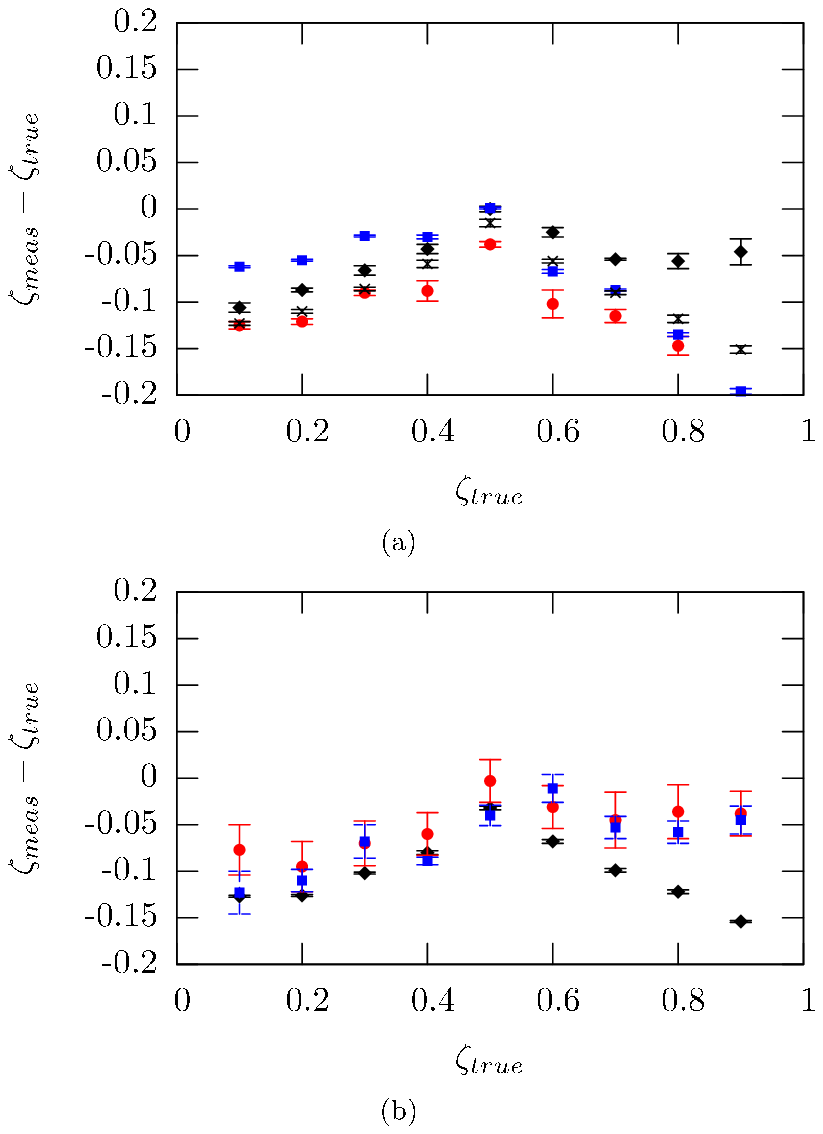}
   \caption{\label{fig:voss0}(Color on-line) $\zeta_{measured} - \zeta_{true}$
for $h_0(x)$ for profiles generated with the Voss algorithm. $L = 512$. a) The 
detrended fluctuation analysis ($\blacklozenge$), the variable bandwidth 
($\bullet$), the Max-Min ($\blacksquare$) and the standard deviation of 
$\Delta h(l)$ ($+$). b) Second order correlation function ($\blacklozenge$), 
power spectrum density analysis ($\bullet$) and averaged wavelet coefficients 
($\blacksquare$).}
   \end{figure}
\end{center}

We conclude this section by checking that the roughness exponent we measure 
from the profiles generated by the Voss algorithm that the  self-affine scaling
originates from the long range correlations in the sign changes. From the 
original profiles $h(x)$ we construct the profile $h_0(x)$ where all jumps are 
on equal size. When $\zeta > 0.5$ one sees from  Fig.\ \ref{fig:voss0} the same
general results as presented above for $h(x)$. The power spectrum analysis, the
averaged wavelet coefficients method and the detrended fluctuation analysis 
give the best estimates. However all estimates are from these three methods are
now systematically $0.05$ to low compared to the true value. Also the global 
methods give good estimates, but these are also to low. The other local methods
show the same large errors as they did for $h(x)$. 

When $\zeta < 0.5$ the picture is is not that clear. The results for all the
different methods show that they measure values above $\zeta_h$, but below 
$1/2$ showing that not all the self-affine information for $\zeta < 0.5$ is
in the sign change correlations for the self-affine profiles generated with
the Voss algorithm.

\section{Size dependence}
\label{sec:sizedep}
\begin{figure}[tbp!]
 \includegraphics[scale = 1.0]{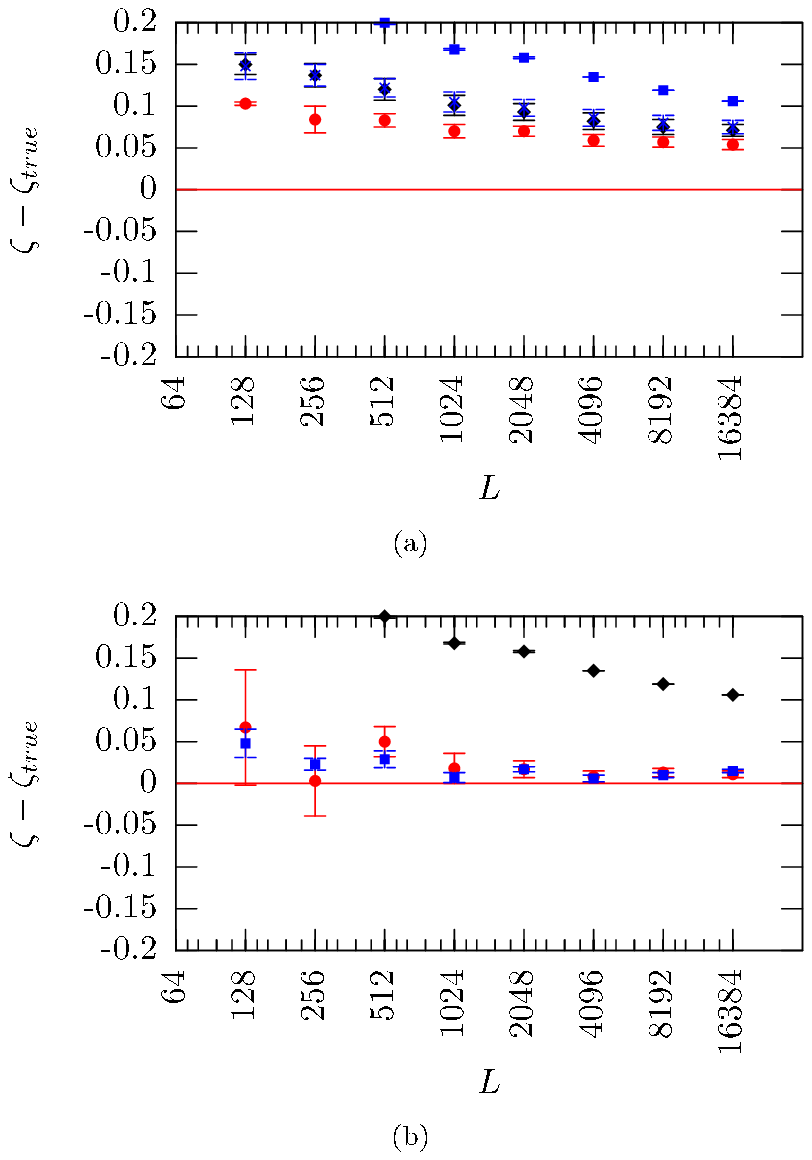}
 \caption{\label{fig:size0.1}(Color on-line) Systematic errors in roughness 
exponent measurements versus system size for $\zeta_{true} = 0.1$. 
a) The detrended fluctuation analysis ($\blacklozenge$), the variable bandwidth
($\bullet$), the Max-Min ($\blacksquare$) and the standard deviation of 
$\Delta h(l)$ ($\times$). b) Second order correlation function ($\blacklozenge$), 
power spectrum density analysis ($\bullet$) and averaged wavelet 
coefficients ($\blacksquare$).}
\end{figure}

\begin{figure}[tbp!]
 \includegraphics[scale = 1.0]{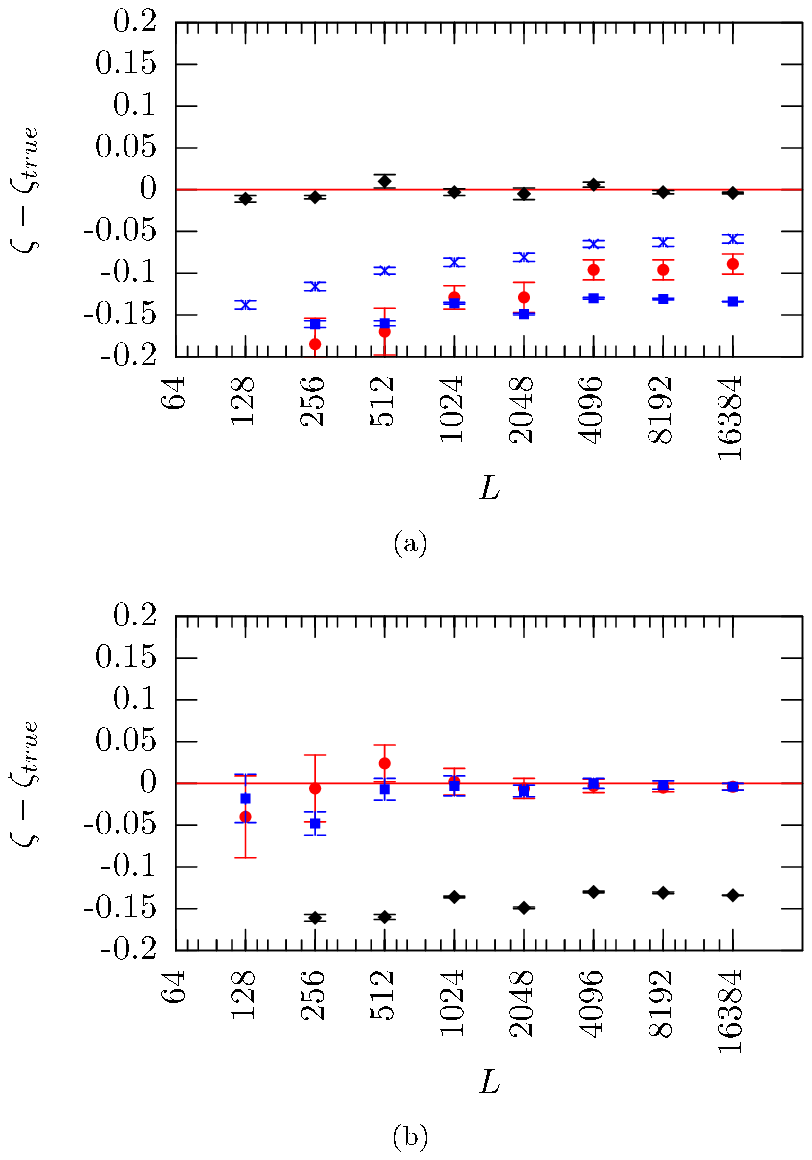}
 \caption{\label{fig:size0.9}(Color on-line) Systematic errors in roughness 
exponent measurements versus system size for $\zeta_{true} = 0.9$. 
a) The detrended fluctuation analysis ($\blacklozenge$), the variable bandwidth
($\bullet$), the Max-Min ($\blacksquare$) and the standard deviation of 
$\Delta h(l)$ ($\times$). b) Second order correlation function ($\blacklozenge$), 
power spectrum density analysis ($\bullet$) and averaged wavelet 
coefficients ($\blacksquare$).}
\end{figure}

In the last section we described how the different measurement methods behaved 
for profiles with $L = 512$. In this section we will expand our study by 
considering how the roughness exponent estimates change with the system size.
The accuracy with which one can measure the roughness exponent is dependent of 
the length of profiles. In \cite{svr95} Schmittbuhl et al.\ reported on the 
size dependence of the error bars for the variable bandwidth method, the 
MAX-MIN method and the power spectrum analysis. We have performed the 
measurements of the roughness exponent with the methods described above on 
samples generated by the Voss algorithm for different system sizes. The results
are presented in Fig.\ \ref{fig:size0.1} to Fig.\ \ref{fig:size0.9}.

As the system size increase the error decreases for the all the measurement 
methods. But even for large system sizes the different methods measure the 
$\zeta$ with systematic errors which are dependent on the value of the $\zeta$.
These systematic errors are similar to the ones shown in Fig. \ref{fig:voss}.

The detrended fluctuation analysis, the power spectrum density analysis and the
averaged wavelet coefficient method all have systematic errors smaller that 
$0.05$ for $L \geq 256$ for all values of $\zeta$. The local window methods 
and the second order correlation function method over-estimate $\zeta$ for 
$\zeta < 0.5$, and under-estimate $\zeta$ for $\zeta \geq 0.5$ as reported in 
Sec.\ \ref{sec:acc}.
\clearpage
\begin{figure}[tbp!]
 \includegraphics[scale = 1.0]{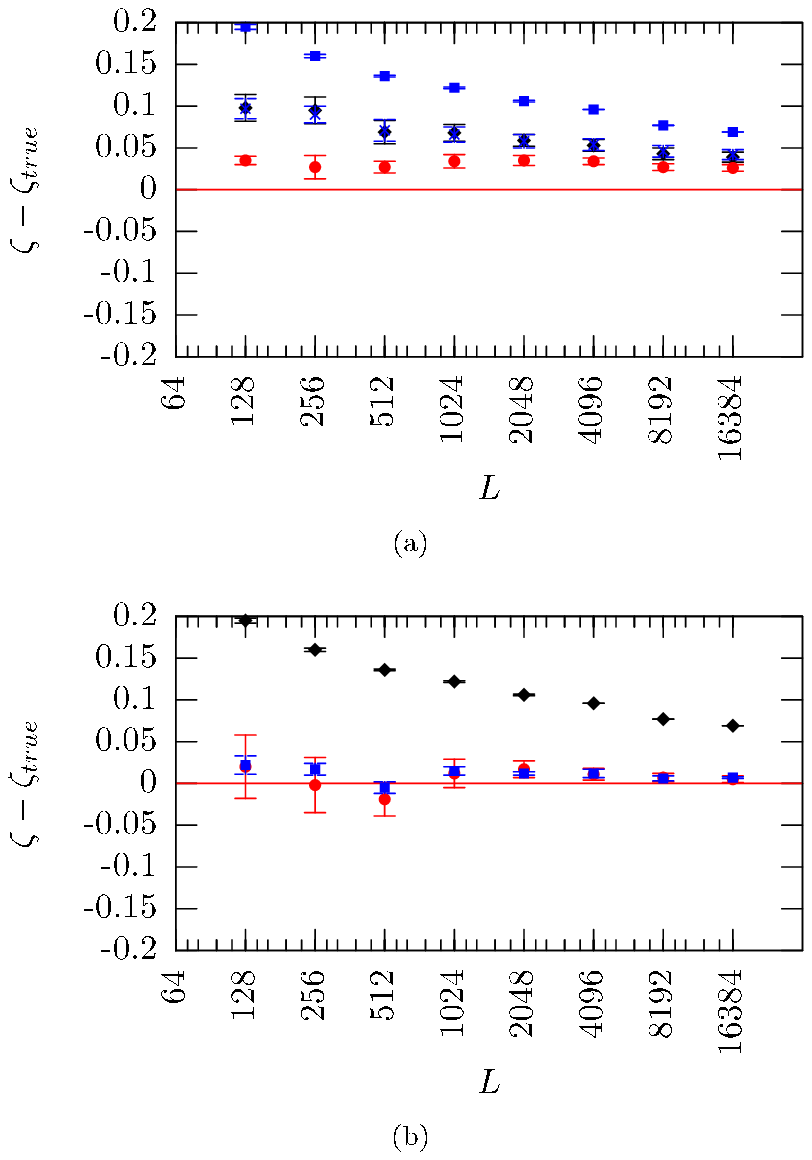}
 \caption{\label{fig:size0.2}(Color on-line) Systematic errors in roughness 
exponent measurements versus system size for $\zeta_{true} = 0.2$. 
a) The detrended fluctuation analysis ($\blacklozenge$), the variable bandwidth
($\bullet$), the Max-Min ($\blacksquare$) and the standard deviation of 
$\Delta h(l)$ ($\times$). b) Second order correlation function ($\blacklozenge$), 
power spectrum density analysis ($\bullet$) and averaged wavelet 
coefficients ($\blacksquare$).}
\end{figure}
\clearpage
\begin{figure}[tbp!]
 \includegraphics[scale = 1.0]{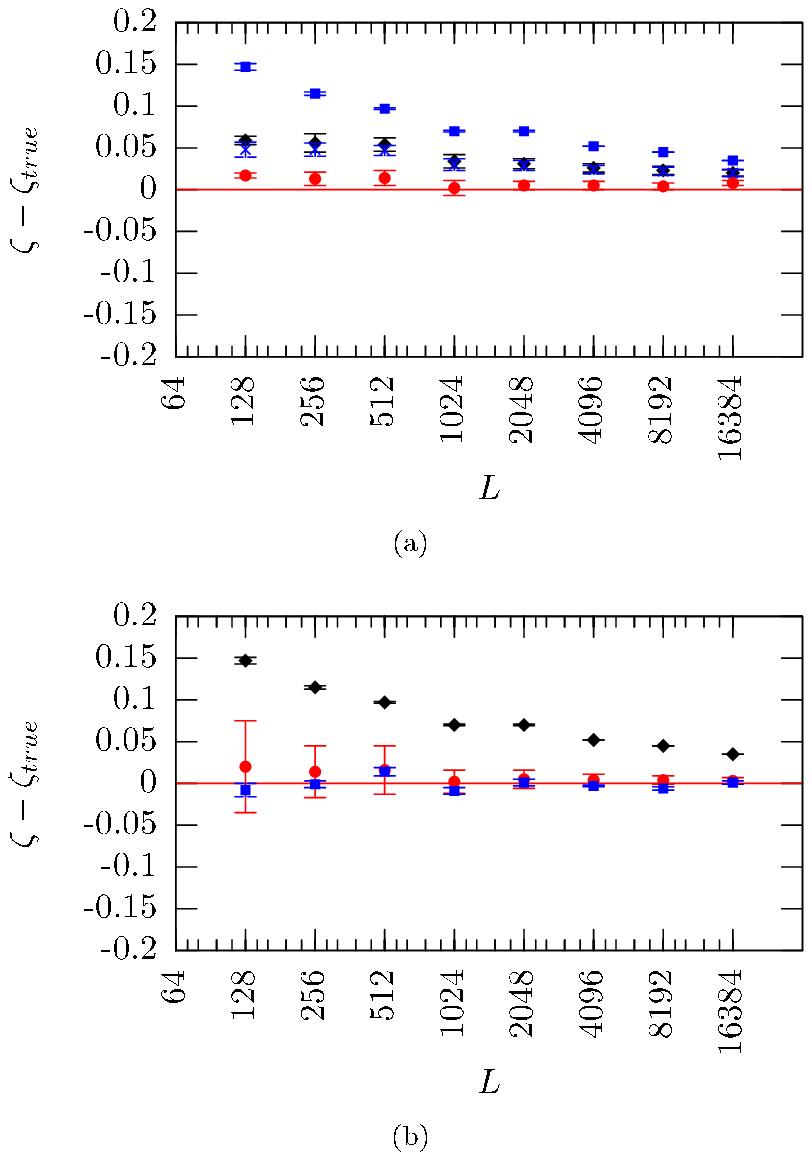}
 \caption{\label{fig:size0.3}(Color on-line) Systematic errors in roughness 
exponent measurements versus system size for $\zeta_{true} = 0.3$. 
a) The detrended fluctuation analysis ($\blacklozenge$), the variable bandwidth
($\bullet$), the Max-Min ($\blacksquare$) and the standard deviation of 
$\Delta h(l)$ ($\times$). b) Second order correlation function ($\blacklozenge$), 
power spectrum density analysis ($\bullet$) and averaged wavelet 
coefficients ($\blacksquare$).}
\end{figure}
\begin{figure}[tbp!]
 \includegraphics[scale = 1.0]{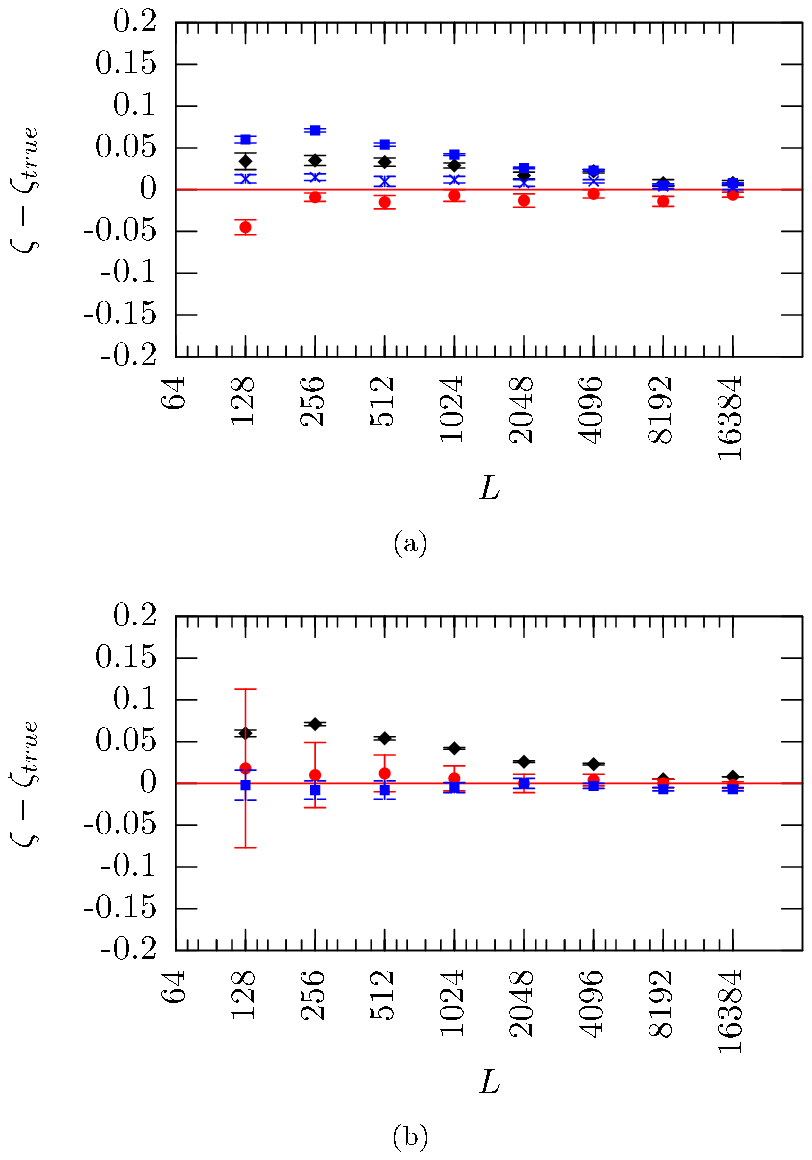}
 \caption{\label{fig:size0.4}(Color on-line) Systematic errors in roughness 
exponent measurements versus system size for $\zeta_{true} = 0.4$. 
a) The detrended fluctuation analysis ($\blacklozenge$), the variable bandwidth
($\bullet$), the Max-Min ($\blacksquare$) and the standard deviation of 
$\Delta h(l)$ ($\times$). b) Second order correlation function ($\blacklozenge$), 
power spectrum density analysis ($\bullet$) and averaged wavelet 
coefficients ($\blacksquare$).}
\end{figure}
\clearpage
\begin{figure}[tbp!]
 \includegraphics[scale = 1.0]{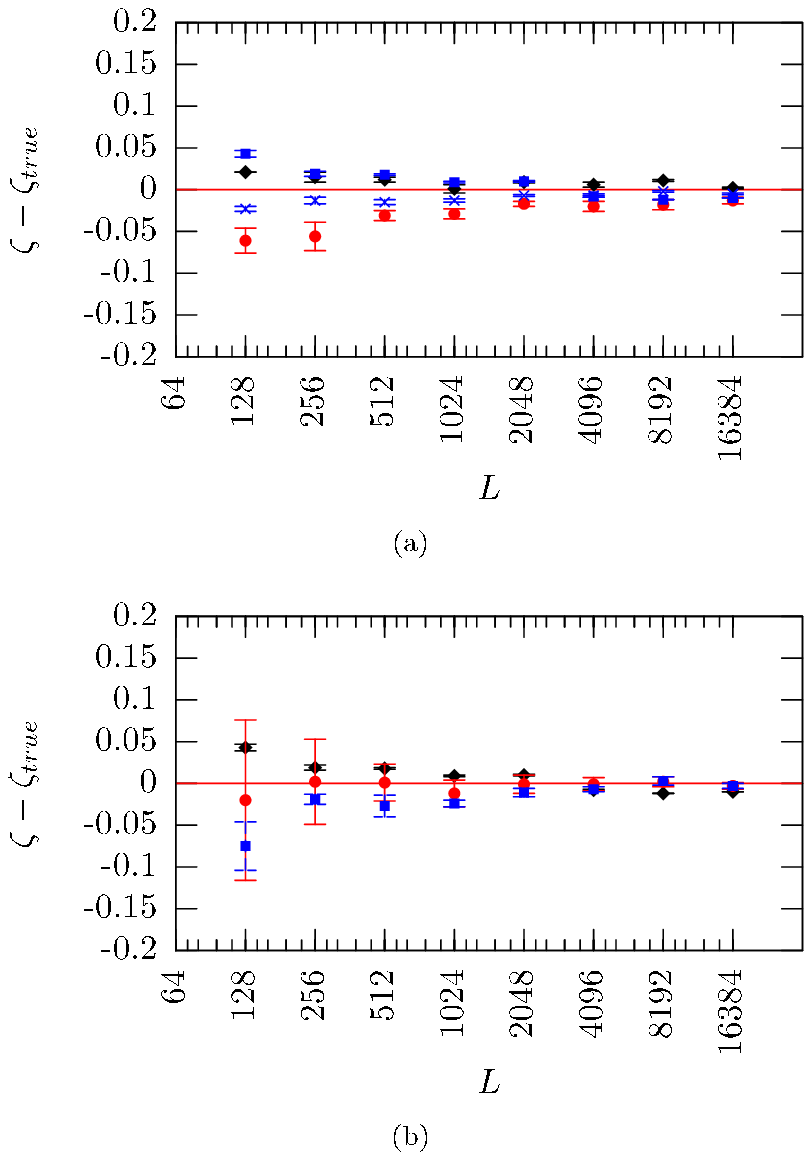}
\caption{\label{fig:size0.5}(Color on-line) Systematic errors in roughness 
exponent measurements versus system size for $\zeta_{true} = 0.5$. 
a) The detrended fluctuation analysis ($\blacklozenge$), the variable bandwidth
($\bullet$), the Max-Min ($\blacksquare$) and the standard deviation of 
$\Delta h(l)$ ($\times$). b) Second order correlation function ($\blacklozenge$), 
power spectrum density analysis ($\bullet$) and averaged wavelet 
coefficients ($\blacksquare$).}
\end{figure}
\begin{figure}[tbp!]
 \includegraphics[scale = 1.0]{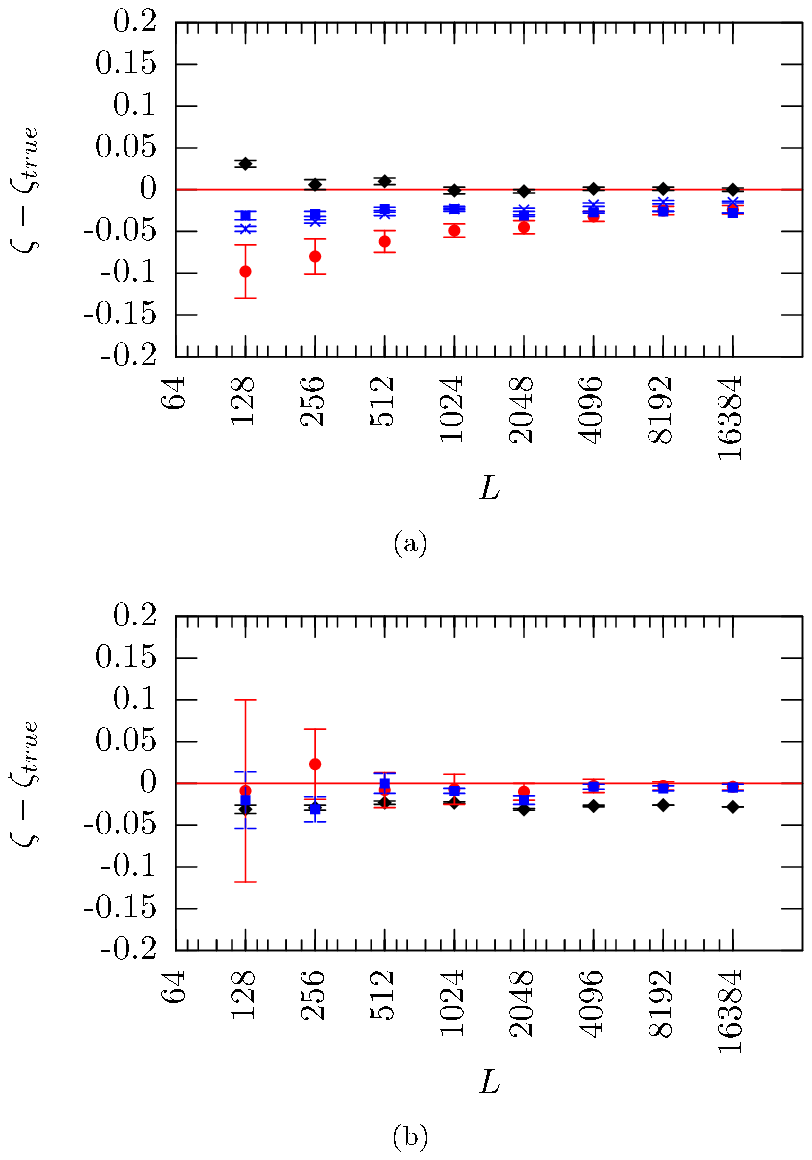}
 \caption{\label{fig:size0.6}(Color on-line) Systematic errors in roughness 
exponent measurements versus system size for $\zeta_{true} = 0.6$.
a) The detrended fluctuation analysis ($\blacklozenge$), the variable bandwidth
($\bullet$), the Max-Min ($\blacksquare$) and the standard deviation of 
$\Delta h(l)$ ($\times$). b) Second order correlation function ($\blacklozenge$), 
power spectrum density analysis ($\bullet$) and averaged wavelet 
coefficients ($\blacksquare$).}
\end{figure}
\clearpage
\begin{figure}[tbp!]
 \includegraphics[scale = 1.0]{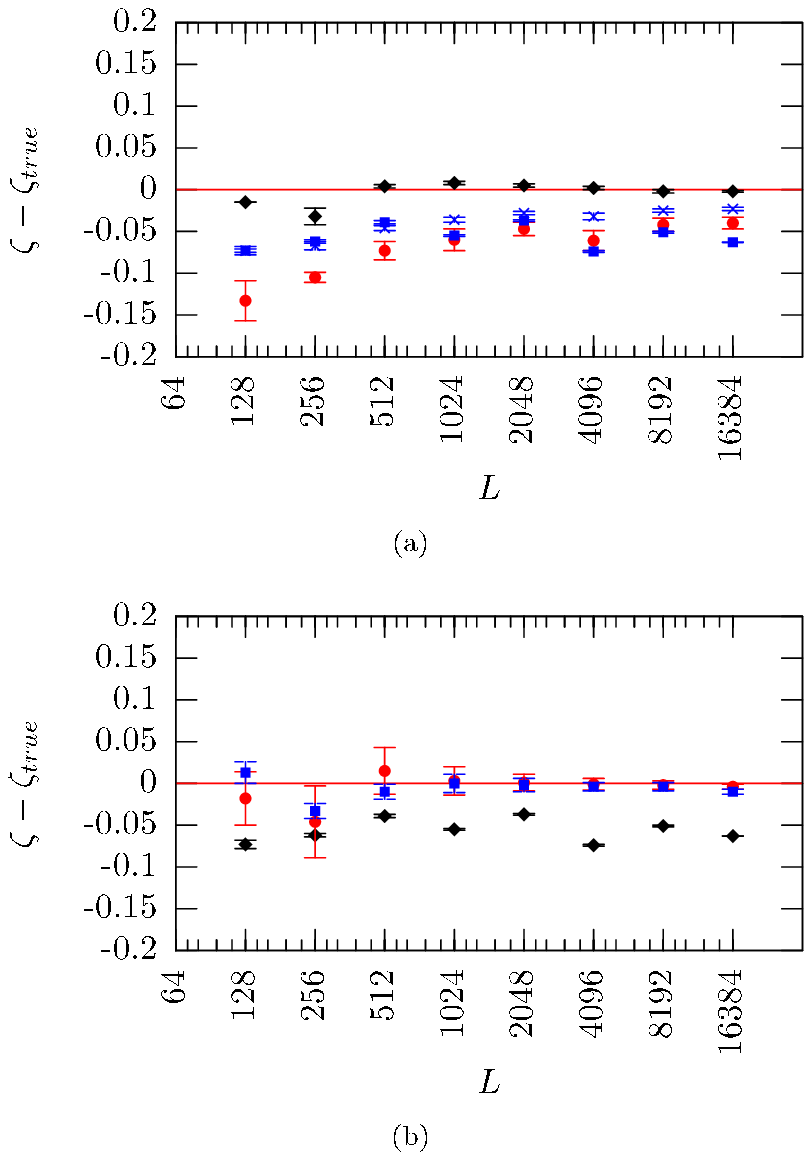}
 \caption{\label{fig:size0.7}(Color on-line) Systematic errors in roughness 
exponent measurements versus system size for $\zeta_{true} = 0.7$. 
a) The detrended fluctuation analysis ($\blacklozenge$), the variable bandwidth
($\bullet$), the Max-Min ($\blacksquare$) and the standard deviation of 
$\Delta h(l)$ ($\times$). b) Second order correlation function ($\blacklozenge$), 
power spectrum density analysis ($\bullet$) and averaged wavelet 
coefficients ($\blacksquare$).}
\end{figure}
\begin{figure}[tbp!]
 \includegraphics[scale = 1.0]{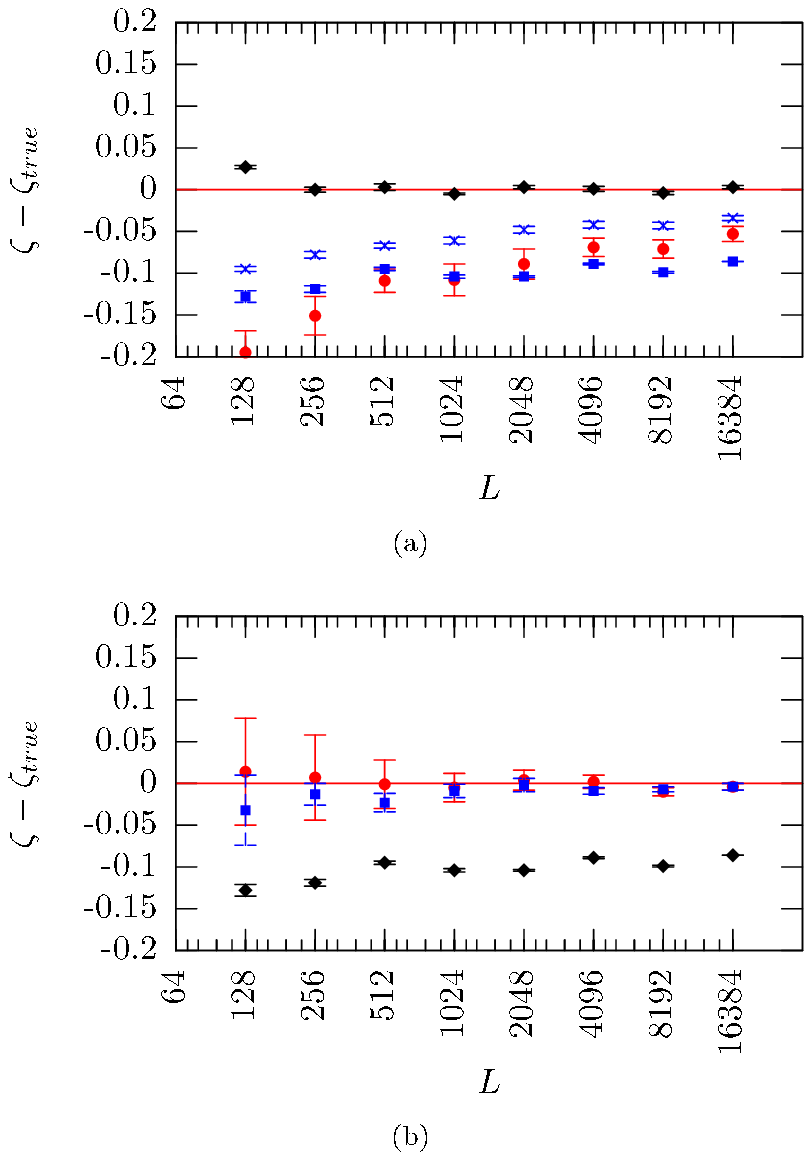}
 \caption{\label{fig:size0.8}(Color on-line) Systematic errors in roughness 
exponent measurements versus system size for $\zeta_{true} = 0.8$. 
a) The detrended fluctuation analysis ($\blacklozenge$), the variable bandwidth
($\bullet$), the Max-Min ($\blacksquare$) and the standard deviation of 
$\Delta h(l)$ ($\times$). b) Second order correlation function ($\blacklozenge$), 
power spectrum density analysis ($\bullet$) and averaged wavelet 
coefficients ($\blacksquare$).}
\end{figure}
\clearpage
\clearpage

\section{Power law distributed step sizes}
\label{sec:powlaw}
\begin{figure}[tbp]
 \includegraphics[scale = 1.0]{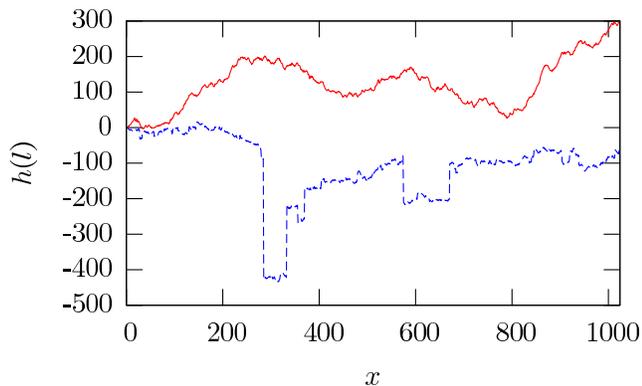}
 \caption{\label{fig:levysaprofiles}(Color on-line) Two samples profiles of 
length $1024$, the one  with solid lines are made with the Voss algorithm with 
$\zeta_{true} = 0.7$, and the dashed one is a random walk with 
L{\'e}vy-distributed jumps with $\alpha = 1.5$}
\end{figure}

\begin{figure}[tbp]
 \includegraphics[scale = 1.0]{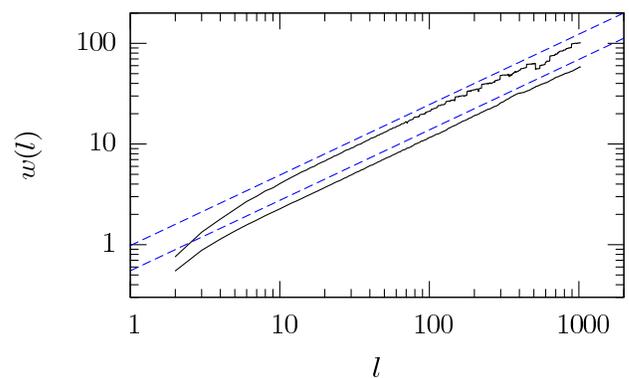}
 \caption{\label{fig:levysadfa}(Color on-line) Roughness exponent measured 
with detrended fluctuation analysis for a profile made with the Voss algorithm 
with $\zeta_{true} = 0.7$(top) and from a random walk with L{\'e}vy-distributed
jumps with $\alpha = 1.5$(bottom).Both the straight dashed lines in the figure 
show $\zeta = 0.7$. 100 samples of length for each type of profiles were used 
in the calculations of the roughness exponent.}
\end{figure}

We will in this section compare two different sets of profiles which have the
same roughness exponent when measured with the detrended fluctuation analysis, 
but different when measured with the power spectrum density analysis. In 
Fig.\ \ref{fig:levysaprofiles} one sample profile from each set of profiles
are shown. One profile made with the Voss algorithm with $\zeta_{true} = 0.7$,
and one profile which is a random walk with  L{\'e}vy-distributed jumps with
$\alpha = 1.5$ are shown. As seen in Fig.\ \ref{fig:levysaprofiles} the main
difference between these two profiles are the existence of the large jumps in 
the second profile. When one use the detrended fluctuation analysis to measure 
the roughness exponent for these different sets of profiles it gives the same 
roughness exponent for both sets as shown in Fig.\ \ref{fig:levysadfa}. The 
only difference is that the amplitude of $w_{DFA}(l)$ is larger for the 
L{\'e}vy-flight profiles than for the Voss profiles. If one use the power 
spectrum analysis, one will measure two different roughness exponents. 
$\zeta = 0.7$ for the profiles made with the Voss algorithm, and $\zeta = 0.5$ 
for the L{\'e}vy profiles.
\begin{figure}[tbp]
 \includegraphics[scale = 1.0]{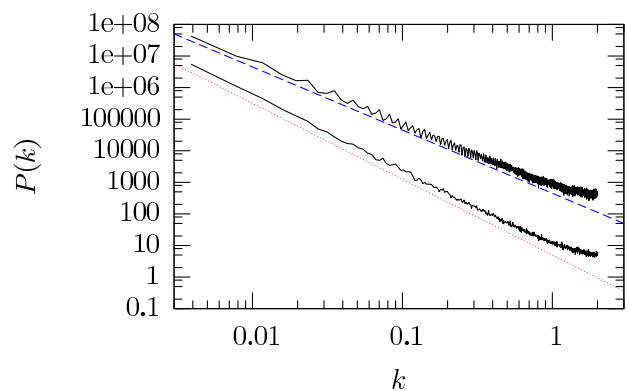}
 \caption{\label{fig:levysapsd}(Color on-line) Roughness exponent measured with PSD for a 
profile made with the Voss algorithm with $\zeta_{true} = 0.7$(bottom) and from
a random walk with L{\'e}vy-distributed jumps with $\alpha = 1.5$(top). The 
corresponding straight lines are for $\zeta = 0.5$ (top) and $\zeta = 0.7$ 
(bottom). 100 samples of length for each type of profiles were used in the 
calculations of the roughness exponent.}
\end{figure}
By using the two modified profiles $h_0(x)$ and $h_r(x)$ described in 
Sec.\ \ref{sec:meas} and measuring the roughness exponent on these one can 
distinguish between the two different sets of profiles. From 
Table \ref{tab:dfas} one sees that the power spectrum density analysis do
not measure $\zeta \neq 0.5$ for the random walk with L{\'e}vy-distributed 
jumps.  Detrended fluctuation analysis will measure $\zeta = 0.7$ except
when the jumps are all set to the same size. For the profiles generated with
the Voss algorithm both the detrended fluctuation analysis and the power
spectrum density analysis give $\zeta = 0.7$ except for when the long range 
correlations are destroyed in $h_r(x)$. This again show that the roughness 
exponent for $\zeta \geq 0.5$ for the Voss generated profiles comes from
the correlations in the sign change. It also shows that the effective roughness
exponent measured with the detrended fluctuation analysis on the L{\'e}vy 
profiles are caused by the power law noise and not the sign change correlation.
In addition one notes that different measurement methods give completely 
different roughness exponents in this case.
\begin{table}[htbp]
  \caption{\label{tab:dfas}Table of roughness exponent measured with the 
detrended fluctuation analysis and the power spectrum analysis on the
  original profiles ($\zeta$), the modified profiles with equal jump size
($\zeta_0$) and the randomly rearranged profiles ($\zeta_r$).}
  \begin{tabular}{ccccccc}
    \hline
    Method & \multicolumn{3}{c}{DFA} & \multicolumn{3}{c}{PSD}\\
    \hline
    Profile set & $\zeta$ & $\zeta_0$ & $\zeta_r$  & $\zeta$ & $\zeta_0$ & $\zeta_r$\\
    \hline					     			   
    Voss        & 0.7 & 0.7 & 0.5 & 0.7 & 0.7 & 0.5\\
    L{\'e}vy    & 0.7 & 0.5 & 0.7 & 0.5 & 0.5 & 0.5\\
    \hline
  \end{tabular}
\end{table}

\begin{figure}[tbp]
 \includegraphics[scale = 1.0]{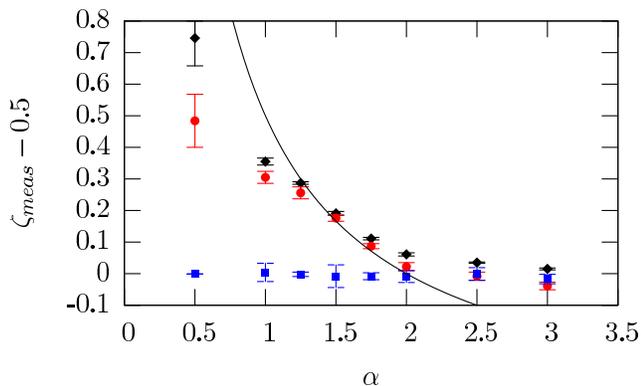}
 \caption{\label{fig:powlaw}(Color on-line) Roughness measurements on random 
walks with power law distributed jumps. The detrended fluctuation analysis 
($\blacklozenge$), the variable bandwidth method ($\bullet$) and the power 
spectrum density analysis ($\blacksquare$). The solid line is the roughness
exponent for a L{\'e}vy-flight, $\zeta_{LF} = 1/\alpha$}
\end{figure}

The power law noise give different corrections to the different measuring 
methods describe in Sec.\ \ref{sec:meas} as was shown earlier in this section. 
To quantify these corrections we have done roughness exponent measurements on 
random walks on which we impose a power law jump distribution with 
different exponents. The length of the samples were $1024$ and the number of 
sample were $1000$. For power laws with exponents in the range 
$\alpha \in \{0.5,3.0\}$ we observe corrections for the local window methods 
and the averaged wavelet coefficients method. As seen in Fig.\ \ref{fig:powlaw}
the corrections are less than $0.05$ for $\alpha \geq 2.0$ and increasing for 
smaller values of $\alpha$.
A L{\'e}vy-flight is defined as
\begin{equation}
  h(l) = \sum_{i=1}^{l} \Delta h_i
  \label{eq:levyflight}
\end{equation}
where $\Delta h_i$ are independent increments following the power law 
distribution described above. From Bouchaud and Gorges \cite{bg90} one 
has that
\begin{equation}
  w(l) = \langle |h(l) - \bar{h}(l)|^2 \rangle \propto l^{1/\alpha},
  \label{eq:levywidth}
\end{equation}
which gives a roughness exponent $\zeta_{LF} = \alpha^{-1}$. When 
$\alpha > 1.2$ $\zeta$ measured with the local window methods and the averaged
wavelet coefficient method follow $\zeta_{LF}$ within $0.05$. For 
$\alpha < 1.2$ $\zeta$ can not keep up with the increasing $\zeta_{LF}$, which 
also increases beyond 1. This is as expected for the variable bandwidth and 
the Max-Min methods as they are restricted to measure roughness exponents in 
the range $\zeta = [0,1]$. The power spectrum analysis and the second order 
correlation function however measures the correlations in the sign change and 
will only measure $\zeta = 1/2$ for any $\alpha$ value as the jumps are
uncorrelated.

As seen above two completely different profiles can with some methods give the 
same roughness exponent when there is power law distributed noise in the 
profiles. In addition this power law noise can give multi-affine corrections 
to the self-affine scaling. 

\begin{figure}[tbp]
 \includegraphics[scale = 1.0]{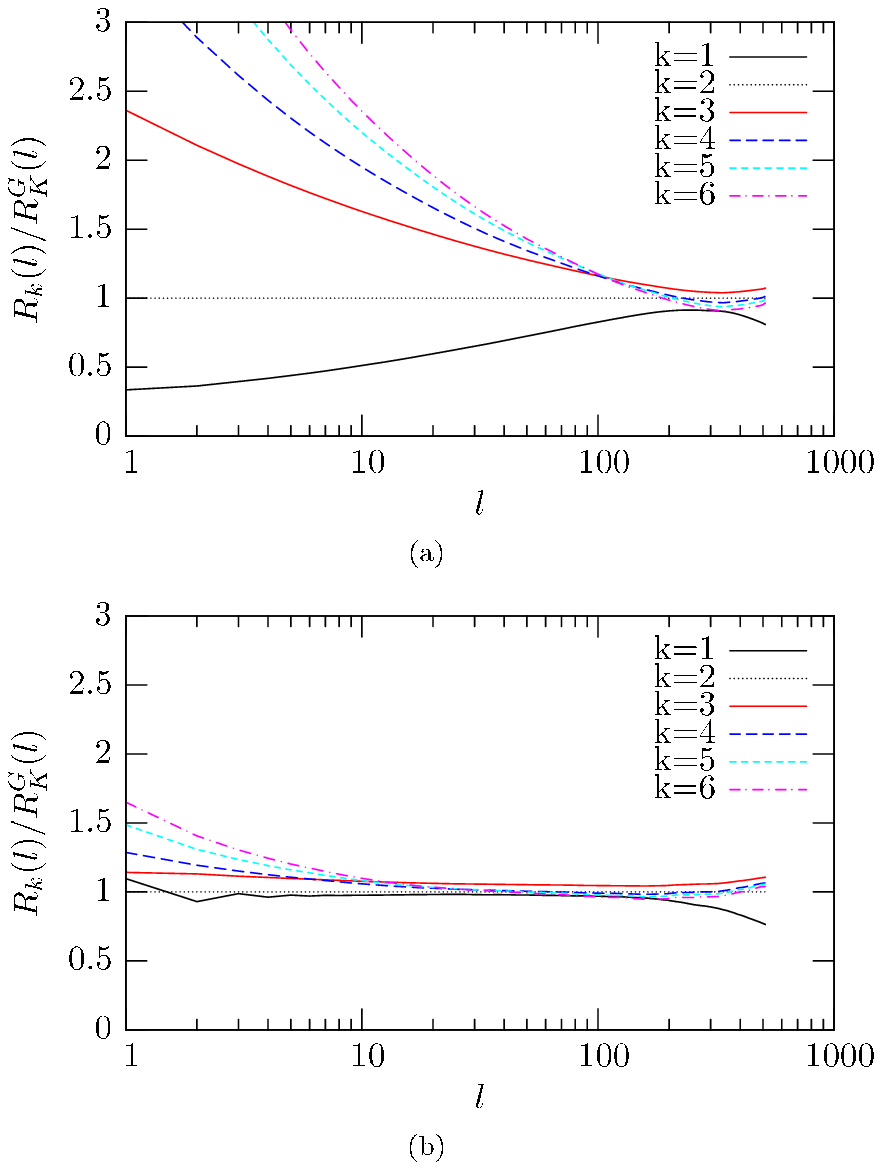}
 \caption{\label{fig:rkapowlaw}$R_k/R_k^G$ for random walks of length $1024$ 
with a power law step size distribution with exponent a) $1.5$ and b $3.0$.}
\end{figure}

Unfortunately fracture profiles from experiments and simulations often have 
noise which partly or completely disturb the self-affine property of the 
profile. This noise can come from the discretization of the data during 
recording of the profile, the grain or fiber size of the material or from 
overhangs in the fracture front. For numerical fracture models recent results 
show that the overhangs in the fracture front have a power law size 
distribution. \cite{brh07,bh07} In Barab{\'a}si and Stanley \cite{bs95}, 
roughening of a surface with an uncorrelated power law noise is shown to give 
multi-affinity below a correlation length $l_X$ which depends on the strength 
of the power law noise given by the exponent $\alpha$, 
i.e. $p(\Delta h) \propto \Delta h^{-\alpha}$, and in \cite{mitchell05}  
Mitchell shows that discontinuities in a self-affine surface give 
multi-affinity. The random walks with a power law jump distribution a have 
multi-affine behavior on small scales. The region with corrections to scaling 
due to multi-affinity changes with $\alpha$ as seen in 
Fig.\ \ref{fig:rkapowlaw}. Here one sees that the profiles  are multi-affine 
for scales smaller than $20$ for $\alpha = 3.0$ and not self-affine at all for 
$\alpha = 1.5$ 

\section{Summary}
\label{sec:disc}

For the system size we have studied here the power spectrum analysis and the 
averaged wavelet coefficient methods gave the best estimates for the roughness
exponent over the range of roughness exponents studied here. While the averaged
wavelet coefficient method is reported to give more accurate results for a 
smaller number of samples, \cite{shn98} this method is prone to systematic 
errors from power law noise. This also applies to the local window methods. We 
have also seen that using the different methods might give roughness exponents 
that differs with a much as $0.1$ with additional errors. In addition several 
of the methods have systematic errors that varies with the value of the 
roughness exponent.

As previously stated by Schmittbuhl et al.\ \cite{svr95} one should use several
different methods for measuring the roughness exponent. Researchers measuring 
the roughness exponent should also take great care in uncover the noise present
in the surfaces that are to be studied. This to make sure that the measuring 
methods chosen are capable of measuring the roughness exponent properly.

Acknowledgments: We want to thank Ingve Simonsen for enlightening discussions 
on L{\'e}vy-flights.

\bibliographystyle{apsrev}

\cleardoublepage

\end{document}